\documentclass[12pt]{article}
\usepackage{epsfig}
\newcommand{\beqn}{\begin{eqnarray}}
\newcommand{\eeqn}{\end{eqnarray}}

\newcommand{\cZ}{{\cal Z}}
\newcommand{\cC}{{\cal C}}

\newcommand{\dd}{{\mathrm d}}
\newcommand{\dD}{{\mathrm D}}

\newcommand{\eq}[1]{~(\ref{#1})}
\date{}
\begin{document}
\title{Short Range Linear Potential \\
in 3D Lattice Compact QED
\vskip-35mm
\rightline{\small ITEP-TH-45/00}
\vskip 40mm
}
\author{B.L.G.~Bakker${}^a$, M.N.~Chernodub${}^b$ and A.I.~Veselov${}^b$\\
{\small\it $^a$ Department of Physics and Astronomy, Vrije Universiteit,}\\
{\small\it De Boelelaan 1081, NL-1081 HV Amsterdam, The Netherlands}\\
\vspace{\baselineskip}
{\small\it ${}^b$ Institute of Theoretical and
Experimental Physics,}\\
{\small\it B.Cheremushkinskaya 25, Moscow, 117259, Russia}\\
\vspace{4\baselineskip}
}
\maketitle
\thispagestyle{empty}
\begin{abstract}
We study the static potential between electric charges in
the finite temperature three dimensional compact gauge theory
on the lattice. We show that in the deconfinement phase at small 
separations between the charges the potential contains a linearly 
rising piece which goes over into the Coulomb potential as the 
distance between the charges is increased. The linear potential is 
due to the gas of magnetic dipoles which are realized as 
monopole--anti-monopole bound states.  
\end{abstract} 

\section{Introduction}
\label{intro}

The compact abelian gauge theory in three space--time dimensions at
finite temperature has two phases separated by the phase
transition~\cite{Parga,Sve,3DcQEDT}.  At low temperature the electric
charges are confined, while at high temperature the confinement
disappears.  These properties of the model originate from the
monopole dynamics, the monopoles being topological defects appearing
due to the compactness of the gauge group.

In the low temperature phase the abelian monopoles form a plasma of
magnetic charges. Due to the long--range nature of the gauge
fields associated with the monopoles the behavior of the plasma is
Coulombic. In the weak coupling regime the plasma is sufficiently
dilute to warrant the use of mean field methods in order to analyse
the non--perturbative behaviour of the system. As it was shown in
Ref.~\cite{Polyakov} test particles of opposite electric charges
being immersed in the magnetic Coulomb plasma experience confining
forces at large separations. The confinement appears because of the
formation of a string--like object between the charges. The string
has a finite thickness of the order of the inverse Debye mass and a
finite energy per unit of string length (``string tension'').

At high temperature the physics of monopoles changes: mo\-no\-po\-les
and anti--mo\-no\-po\-les form magnetically neutral bound
states~\cite{AgasianZarembo}. These states are filling up the
vacuum with a neutral dipole plasma and the confinement of electric
charges disappears. This phenomenon can be explained in two ways.
First, the field of the magnetic dipoles is much weaker at large
distances compared to the field of the monopoles and thus the dipole
is unable to induce a non--zero string tension.  The second
explanation is that in the dilute dipole plasma the Debye screening
is absent~\cite{no-screening} and effectively this corresponds to an
infinitely thick confining string. Since the electric flux of the
string is a conserved quantity, it is no more concentrated in a small
region (``core of the string'') and thus the string ``dissolves'' as
we go from low to high
temperatures\footnote{Another mechanism of
the phase transition in the Georgi--Glashow model due to magnetic
vortex dynamics is discussed in Ref.~\cite{Alex}.}.

Despite the fact that the dipoles themselves are unable to create the
confining string, they may still be responsible for the
non--perturbative physics of electric charges. The
string tension corresponding to the linear part of the long
distance potential of electric charges in a pure magnetic
monopole gas increases with addition of the magnetic dipoles
to the monopole gas~\cite{Ch00-2}.
At short distances the potential of electric charges
in a pure magnetic dipole gas contains a piece which rises
linearly with the distance between the charges~\cite{Ch00-1}.

Note that nontrivialities in a short distance potential appear also
in the zero--tem\-pe\-ra\-tu\-re theory. According to
Ref.~\cite{ChGuPoZa:GG}, at distances $R$ much smaller than the
correlation length of the monopole--anti-monopole plasma the
potential contains a perturbative contribution due to one--photon
exchange plus a non--perturbative piece proportional to $R^\alpha$,
$\alpha \approx 0.6$.  A non--trivial short-distance potential may
have many physically interesting consequences, see, {\it e.g.}
Ref.~\cite{VIZandCo} for a discussion in the context of QCD and other
theories.
The monopole binding in 3D compact QED is qualitatively
similar to the formation of the instanton molecules in the high
temperature phase of QCD suggested to be responsible for the chiral
phase transition~\cite{Shuryak}.

The structure of this paper is as follows. In the next section we
review some results concerning the physics of the dipole gas in the
continuum following Ref.~\cite{Ch00-1}. Sect.~\ref{results}
contains the results of our simulation of the compact abelian gauge
model in three dimensions. We show that indeed the short-distance
potential contains a linear piece, which can be explained as due to the
non--trivial dynamics of the gas of lattice dipoles. Our conclusions
are summarized in the last section.

\section{Dipole gas in continuum}
\label{overview}

A magnetic dipole is a magnetically neutral localized pair of a
monopole and an anti-monopole separated by a distance $r$. The magnetic
moment of the dipole is $\vec\mu = g_m \, \vec r$ where $g_m$ is the
magnetic charge of the constituent monopole and $\vec r$ is the
relative position vector pointing from the anti-monopole to the
monopole. If the typical distance between dipoles is much larger than
the dipole size $r$, then the dipoles may be treated as point--like
particles. This condition can be written as follows:
\beqn
 \xi = \rho^{\frac{1}{3}} \, r \ll 1\,,
 \label{diluteness}
\eeqn
where $\rho$ is the mean density of the dipole gas. We shall see below
that in our calculations this condition is always fulfilled.

The action of two interacting point--like dipoles with magnetic
moments $\vec\mu_a$ and $\vec\mu_b$ located at positions $\vec x_a$
and $\vec x_b$, respectively, is given by the formula:
\beqn
 V(\vec\mu_a,\vec\mu_b;\vec x_a,\vec x_b) =
 (\vec \mu_a \cdot \vec \partial)\, (\vec \mu_b \cdot \vec \partial)
 D_{(3D)}(\vec x_a-\vec x_b)\,,
\label{basic:interaction}
\eeqn
where $D_{(3D)}(\vec{x}) = 1/(4 \pi |x|)$ is the propagator for a scalar
massless particle in three dimensions.

Below we consider the case of the fixed absolute values of the dipole
moments, $\mu_a = \mu = {\mathrm{const}}$. The case of fluctuating
dipole moments is considered in Ref.~\cite{Ch00-1}.

The statistical sum of the dilute dipole gas can be written as follows:
\beqn
\cZ = \sum\limits^\infty_{N=0} \frac{\zeta^N}{N!}
\int \dd^3 x_1 \int\dd^3 \mu_1 \cdots
\int \dd^3 x_N \int\dd^3 \mu_N
\exp\Biggl[ - \frac{1}{2} \sum^N_{\stackrel{a,b=1}{a \neq b}}
V(\vec\mu_a,\vec\mu_b;\vec x_a,\vec x_b)\Biggr]\,,
\label{GasPF}
\eeqn
where $\zeta$ is the dipole fugacity. As in the case of the monopole
gas~\cite{Polyakov} the dipole fugacity is a non--perturbative quantity,
since $\zeta \sim e^{-S_0}$, where $S_0 \sim g^{-2}_e$ is
the action of a single dipole  and $g_e = 2 \pi \slash g_m$ is the
fundamental electric charge in the theory.

The partition function~\eq{GasPF} can be rewritten as
follows~\cite{Ch00-1},
\beqn
\cZ = \int \dD \chi \, \exp\Bigl\{ - \int \dd^3 x \,
\Bigl[\frac{1}{2} {(\vec \partial \chi)}^2 - 4 \pi \zeta \,
\frac{\sin (\mu |\vec \partial \chi|)}{\mu |\vec \partial \chi|}\,
\Bigr]\Bigr\}\,,
\label{PF:basic}
\eeqn
where $|\vec \partial \chi| = \sqrt{{\vec \partial \chi}^2}$.
Rescaling the field,
$\chi \to {(4 \pi \zeta)^{- 1 \slash 3}} \, \mu ^{-1} \, \chi$,
and the coordinates,
$x \to {(4 \pi \zeta)^{- 1 \slash 3}} x$,
we immediately realize
that the dynamics of the gas is controlled by the dimensionless
constant
\beqn
\lambda=4 \pi \zeta \mu^2\,.
\label{lambda}
\eeqn

The vacuum expectation value of the dipole density,
$\rho_d(x) = \sum\nolimits_i \delta^{(3)} (x - x_i)$,
is
\beqn
 \rho_d \equiv <\rho_d(x)> = 4 \pi \zeta \Bigl(1 + O(\lambda)\Bigr)\,,
 \label{dens:quant}
\eeqn
where the last equality is valid for small couplings $\lambda$.  In
this regime the coupling $\lambda$ is proportional to the  density of
the dipoles, $\lambda = \rho_d \mu^2$, up to $O(\lambda^2)$ terms.
Therefore the smallness of the parameter $\lambda$ can be interpreted
as a requirement for the density of the dipole moments to be small:
\beqn
\rho_d \, \mu^2 \ll 1\,.
\label{lambda:mu}
\eeqn

The static interaction of particles with electric charges $g_e$ is the
sum of the perturbative contribution from the one--photon exchange and
the contribution from the dipole gas, respectively:
\beqn
 V(R) = \frac{g^2_e}{2}\, D_{(2D)}(R) + E^{\mathrm{gas}}(R)\,,
 \quad E^{\mathrm{gas}}(R) = - \lim\limits_{T \to \infty} \frac{1}{T}
 <W(\cC_{R\times T})>\,,
 \label{full:potential}
\eeqn
where $\cC_{R\times T}$ stands for the rectangular $R\times T$
trajectory of the test particle and $D_{(2D)}(R) = - {(2 \pi)}^{-1}
\, \log(mR)$ is the two--dimensional propagator for a scalar massless
particle, $m$ is a regulator of the dimension of mass. We consider
here test particles with an electric charge equal to one unit of the
fundamental charge. The case of particles of arbitrary charges is
discussed in Ref.~\cite{Ch00-1}.

At small distances between the test electric charges, $R \ll r$, the
behaviour of the potential was shown to be as follows~\cite{Ch00-1}:
\beqn
E^{\mathrm{gas}}_{\mathrm{cl}}(R) = \sigma R \, \Bigr[1 +
O\Bigl(R \slash r \Bigr)\Bigr]\,, \quad
\sigma = \pi^3 \, \zeta \, r =
\frac{\pi^2}{4} \rho_d \, r\,, \quad R \ll r\,.
\label{sigma:th}
\eeqn
where $\sigma$ is given to leading order in $\lambda$.

At large distances, $R \gg r$, the classical energy
of the electric charges in the dipole gas grows logarithmically:
\beqn
E^{\mathrm{gas}}_{\mathrm{cl}}(R) = \frac{8 \pi^2}{3}
\zeta r^2 \log\Bigl(\frac{R}{r}\Bigr)
\, \Bigr[1 + O\Bigl(r \slash R \Bigr)\Bigr]\,,\quad R \gg r\,.
\label{Ecl:long}
\eeqn
According to eq.~\eq{full:potential} the dipole gas
non--perturbatively renormalizes the coupling constant $g_e$ at large
distances and the full potential \eq{full:potential} has the
following behaviour:
\beqn
V(R) = \frac{g^2_e}{2 \epsilon}\, D_{(2D)}(R)\,, \quad
\epsilon = 1 + \frac{1}{3}\, \lambda + O(\lambda^2)\,,\quad
R \gg r\,,
\label{Vfull}
\eeqn
where $\epsilon$ is the dielectric constant (permittivity).

The linear term at small distances \eq{sigma:th} is a result of the
interaction of the dipole clouds which surround the external electric
sources. Indeed, in three dimensions the static test charge
trajectories $j_\cC$ may be considered as electric currents running
along the "wires" $j_\cC$.  These currents induce a magnetic field
which encircles the trajectories and is defined to leading order
by the classical Maxwell equations. The closer to the current the
larger the magnetic field is. Since the energy of a magnetic dipole
becomes lower with increasing magnetic field strength, the density of
the magnetic dipoles should increase towards the position of the
electric test charges. Therefore the dipoles form clouds near test
particles and the interaction of these clouds is responsible for the
non--perturbative part of the inter--particle potential.

\section{Compact QED in 3D: Numerical Results}
\label{results}

We consider the $3D$ compact $U(1)$ gauge model with the standard
Wilson action, $S_P = - \beta \cos \theta_P$, where $\theta_P$ is the
abelian field strength tension constructed from the abelian compact
fields $\theta_l$ and $\beta$ is the lattice coupling which
is related to the electric charge $g_e$ in the following way,
\beqn
\beta = \frac{1}{a g^2_e}\,,
\label{betalat}
\eeqn
$a$ is the lattice spacing. Our numerical results are obtained on the
lattice of size $32^2 \times 8$. The finite temperature phase transition
corresponds to the lattice coupling $\beta_c = 2.3$,
Ref.~\cite{3DcQEDT}. The high--temperature phase corresponds to
$\beta > \beta_c$.

We do realize that our results can not be expected to work in the
confinement phase. Therefore, when fits are made only the numerical
results for $\beta > \beta_c$ are used. Still, we consider it worthwhile
to show the numerical results in a larger domain of $\beta$ values as
they illustrate the qualitative difference between the two regimes.

\begin{figure}[!htb]
 \begin{minipage}{15.0cm}
 \begin{center}
  \epsfig{file=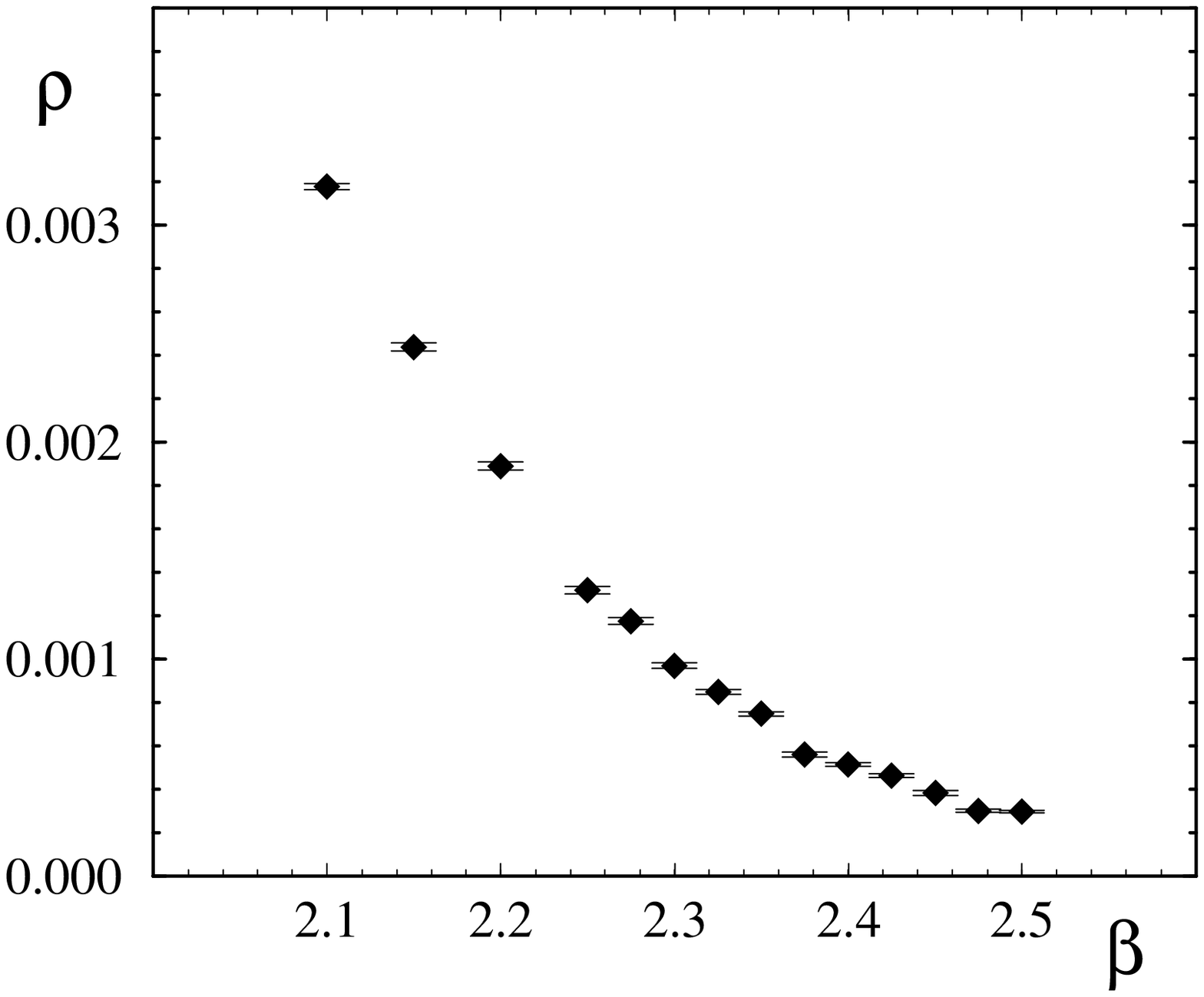,width=7.0cm,height=4.5cm} \hspace{3mm}
  \epsfig{file=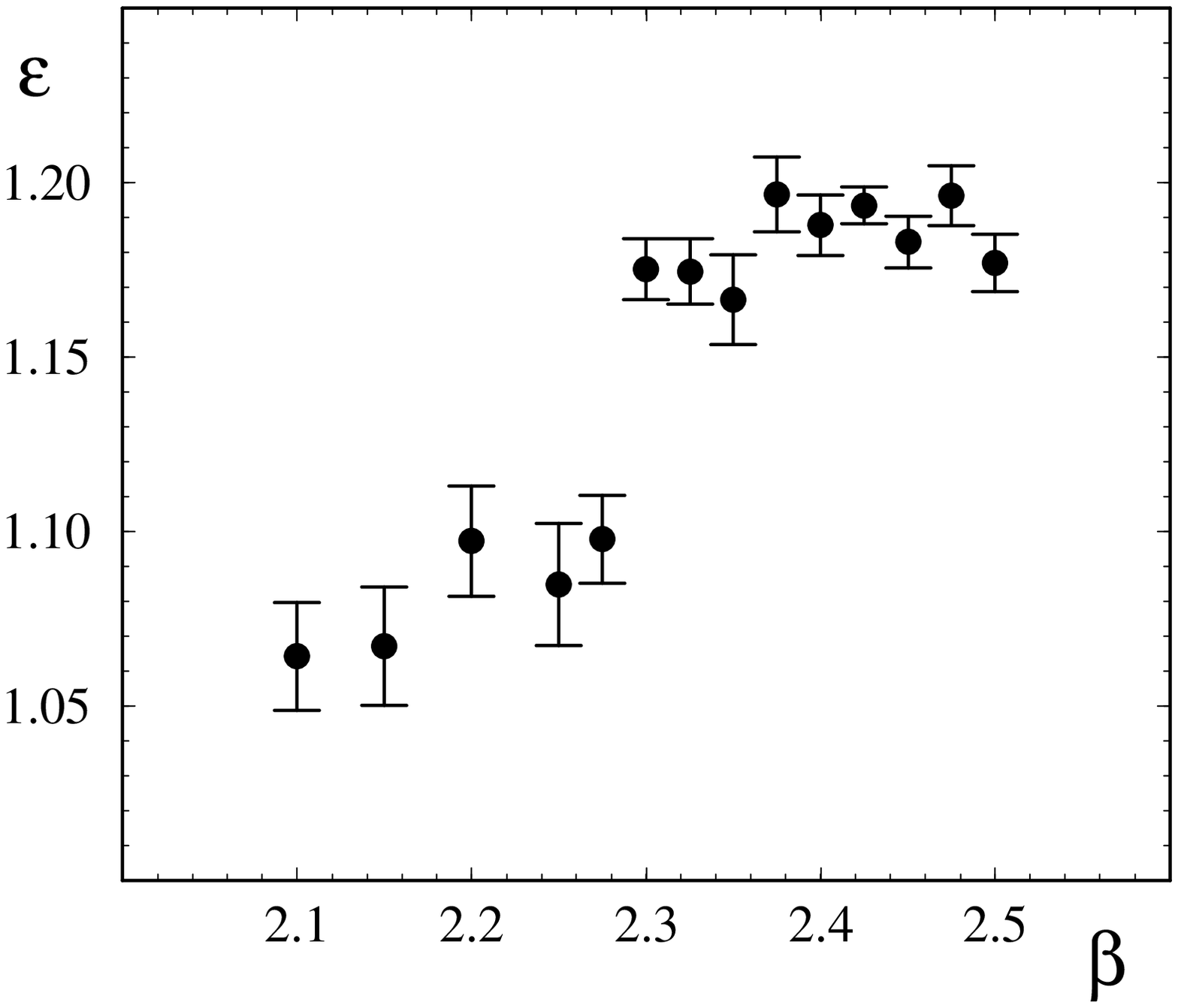,width=7.0cm,height=4.52cm} \\
$(a)$ \hspace{7.0cm} $(b)$\\
 \end{center}
 \end{minipage}
\caption{(a) The density of the abelian monopoles, $\rho$,
$vs.$ lattice coupling $\beta$; (b) vacuum dielectric constant
(permittivity) $\epsilon$.}
\label{figs:dens:epsilon}
\end{figure}

We show the behaviour of the density $\rho$ of the abelian monopoles
$vs.$ $\beta$ in Fig.~\ref{figs:dens:epsilon}(a). The density $\rho$
decreases rapidly with increasing of $\beta$ ($i.e.$, with increasing
of the temperature). Note that in the deconfinement phase, $\beta >
\beta_c$ the density of the monopoles is non--zero.
Analysis of the gauge field configurations gives that the monopoles
appear as tight monopole-anti-monopole bound states in which the
constituents are separated basically by one or two lattice spacings
$a$.

\vspace{5mm}
\vspace{5mm}

We study the potential between oppositely electrically charged
particles measuring the correlations of two Polyakov lines, $L(x)$,
separated by a distance $R$,
\beqn
V(R) = - \log <L(0) L^+ (R)>\,.
\label{Vnum}
\eeqn
The lattice potential $V(R)$ is fitted by the following formula:
\beqn
V(R) = - \frac{1}{\beta_{\mathrm {fit}}} D(R)
- \log\Bigl\{\cosh\Bigl[\sigma\, (R- L \slash2)\Bigr] \Bigr\} + C\,,
\label{Vfit}
\eeqn
where the effective coupling, $\beta_{\rm fit}$, the short string tension,
$\sigma$, and the additive energy renormalization, $C$, are the fitting
parameters. The function $D(R)\equiv D(R,0)$ is the two--dimensional
lattice propagator for a scalar massless particle:
\beqn
 D(\vec x) = \frac{1}{2 L^2} \sum\limits^{L-1}_{
 \stackrel{k_1,k_2=0}{\vec k^2 \neq 0}}
 \frac{e^{2 \pi i \, \vec k \cdot \vec x\slash L}}
 {2 - \cos(2 \pi k_1 \slash L) - \cos(2 \pi k_1 \slash L)}\,,
 \label{Dlat}
\eeqn
where the zero mode, $k_1 = k_2 =0$, is excluded. We are using the
massless propagator since in the high temperature phase (which
we are interested in) the potential between electric charges is long
ranged according to eq.\eq{Vfull} (the dipoles are unable to induce a
finite correlation length). Function \eq{Vfit} fits the numerically
obtained potential \eq{Vnum} with $\bar{\chi}^2 = \chi^2 \slash
d.o.f \sim 1$ up to distances $r a \sim 8$, while at large distances
the function $\bar{\chi}^2$ become unacceptably large. Thus we are
fitting the potential \eq{Vnum} by the function \eq{Vfit} at the
distances $R = 0, \dots, 8$.

The second term in Eq.~\eq{Vfit} corresponds to the linear term
modified by finite volume corrections. In the limit $L\to \infty$
this term transforms to the ordinary linear term $\sigma R$. Note
that the finite volume corrections in the first term are already
taken into account since according to Eq.~\eq{Dlat}, $D(R+L) = D(R)$.

The coupling $\beta_{\mathrm {fit}}$ used as a fitting parameter in
Eq.~\eq{Vfit} should be equal to the ``bare'' coupling, $\beta$, in
first order perturbation theory. However, due to the
non--perturbative  corrections coming from the dipole
gas~\cite{Ch00-1}, this coupling gets
renormalized.
The ratio of the renormalized and bare couplings, according to
Eqs.~(\ref{Vfull},\ref{betalat}) is the dielectric constant
(permittivity):
\beqn
 \epsilon(\beta) = \frac{\beta_{\rm
 fit}(\beta)}{\beta} = \frac{g^2_{e,\mathrm{bare}}}{g^2_e}\,,
 \label{ratio}
\eeqn
which is plotted in
Fig.~\ref{figs:dens:epsilon}(b). In the deconfinement region,
$\beta > \beta_c$,
close to the phase transition
the dielectric constant $\epsilon$ is
approximately equal to $1.2$.
A similar investigation of the charge
renormalisation in four dimensional zero--temperature QED has been
done in Refs.~\cite{charge}.

\begin{figure}
 \begin{minipage}{15.0cm}
 \begin{center}
  \epsfig{file=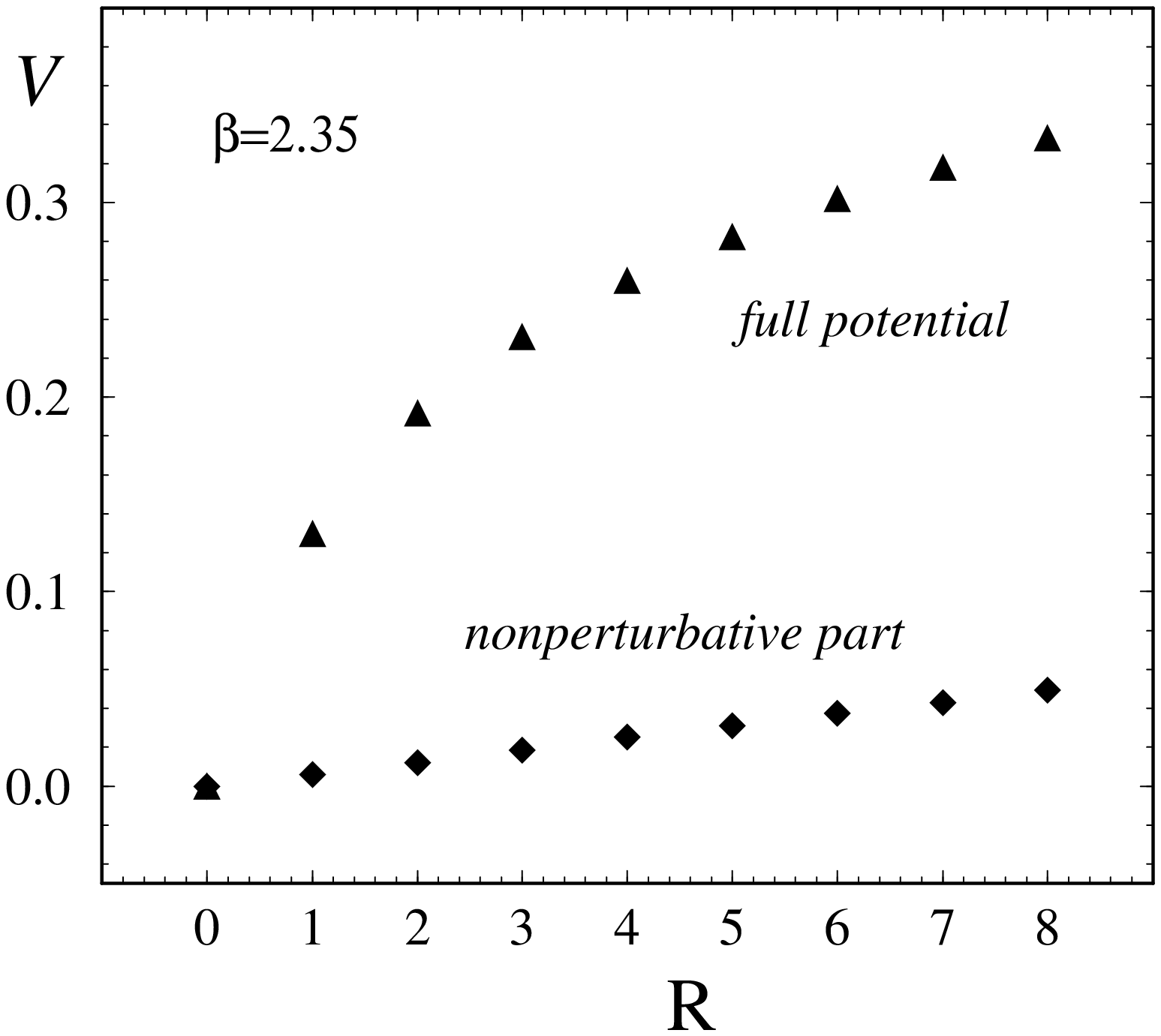,width=7.2cm,height=5.0cm}
  \hspace{3mm}
  \epsfig{file=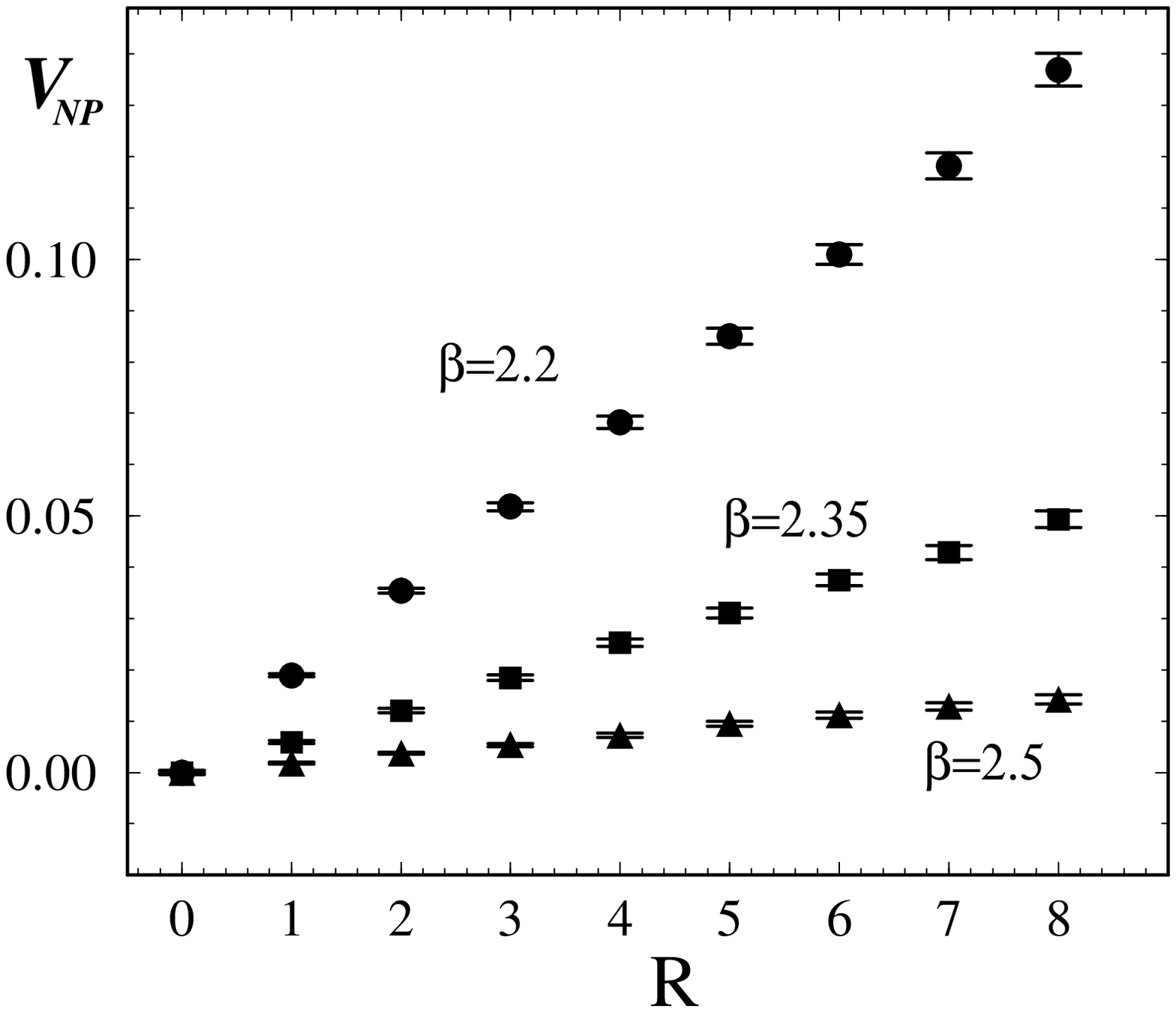,width=7.2cm,height=5.0cm} \\
$(a)$ \hspace{7.0cm} $(b)$ \\
\vspace{4mm}
 \end{center}
 \end{minipage}
\caption{
(a) Nonperturbative  part of the test particle potential,
eq.~\eq{Vnp:eq}, $vs.$ full potential, for $\beta = 2.35$, and (b)
the nonperturbative parts of the potential for various values of
$\beta$.}
\label{figs:v}
\end{figure}

The non--perturbative part of the inter--particle potential is given
by the full potential minus the perturbative one--photon exchange,
\beqn
V_{NP}(R,\beta) = V_{\rm exp}(R,\beta) -
\frac{1}{\beta_{\rm fit}(\beta)} D(R)\,.
\label{Vnp:eq}
\eeqn
In Eq.~\eq{full:potential} the non--perturbative part of the
potential is solely due to the dipole gas contribution,
$E^{\mathrm{gas}}(R)$, and, according to our fits it is linear. We
show the full potential and the non-perturbative part of the
potential $\beta=2.35$ in Fig.~\ref{figs:v}(a).
Indeed, we observe that the non--perturbative part is linear and is
substantially smaller than the perturbative one. The non-perturbative
parts of the potential for various $\beta$ are shown in
Fig.~\ref{figs:v}(b).

\begin{figure*}[!htb]
 \begin{minipage}{15.0cm}
 \begin{center}
  \epsfig{file=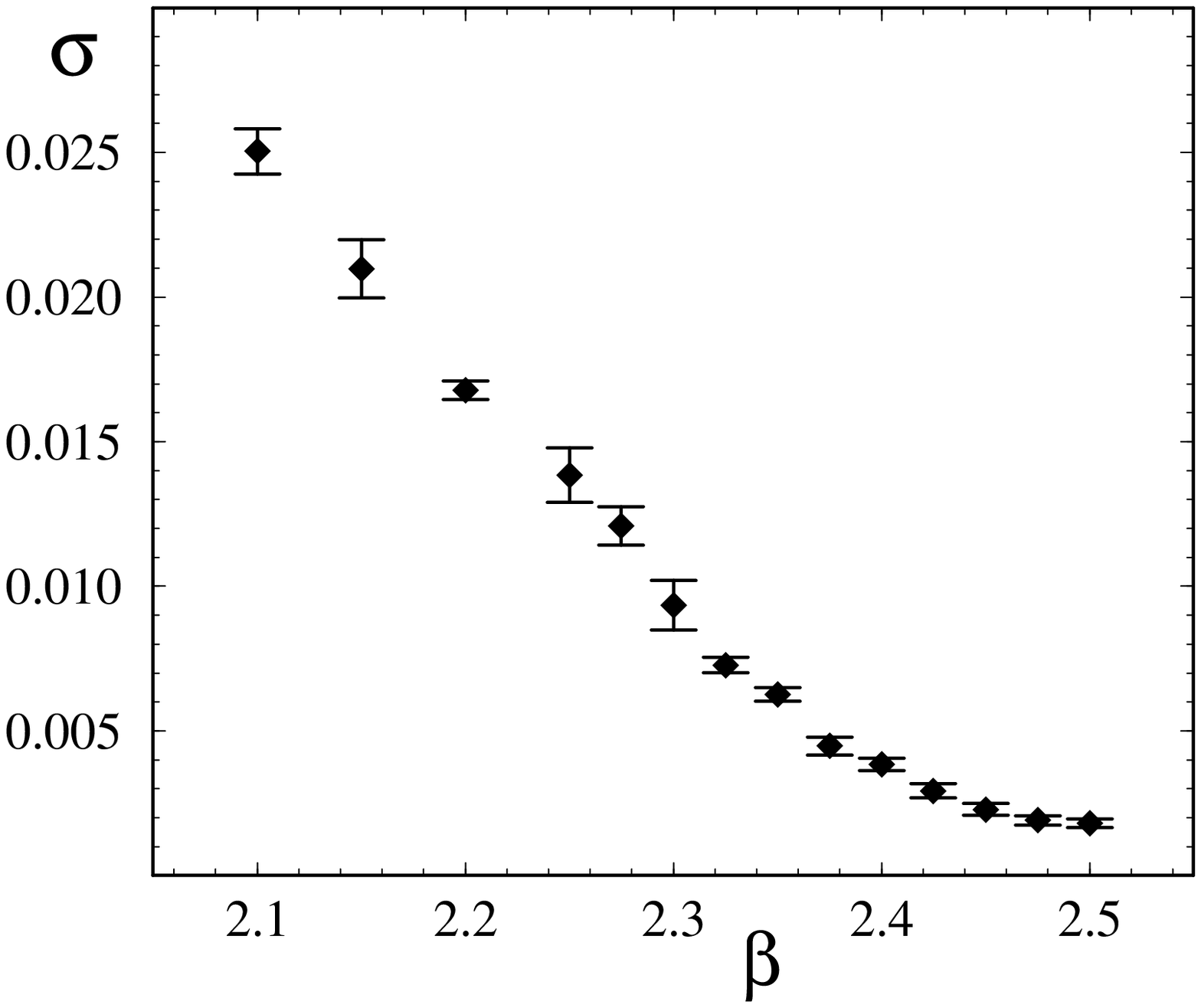,width=7.0cm,height=4.6cm} \hspace{3mm}
  \epsfig{file=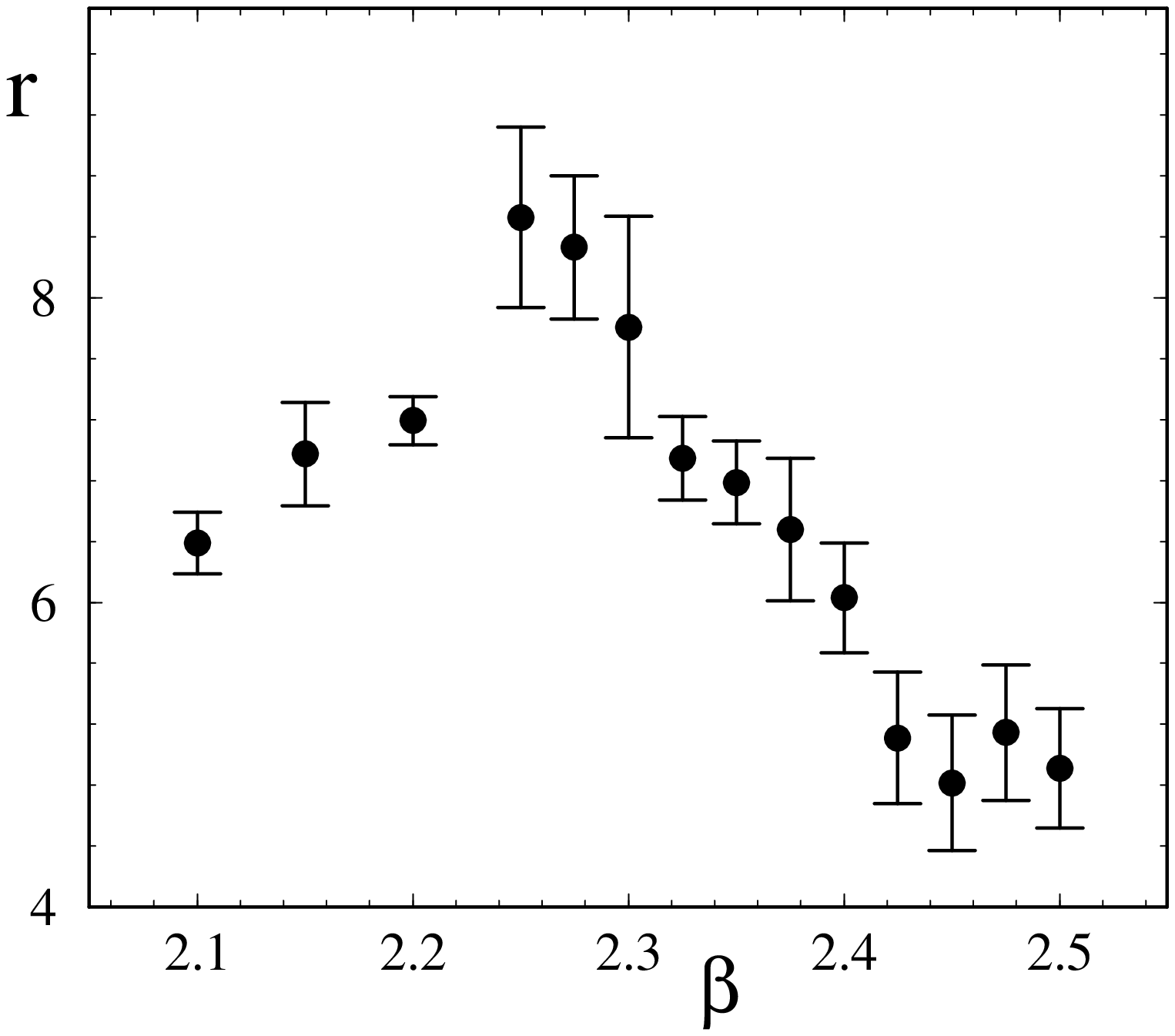,width=7.0cm,height=4.5cm} \\
 $(a)$ \hspace{7.0cm} $(b)$\\
 \end{center}
 \end{minipage}
 \caption{(a) the average distance between constituent monopoles in a
dipole state, eq.~\eq{r}; (b) the short distance string tension
obtained with the help of the fit~\eq{Vfit}.}
\label{figs:r:str}
\end{figure*}

The tension of the short string $\sigma$, defined in eq.\eq{Vfit} is plotted
{\it vs.} $\beta$ in Fig.~\ref{figs:r:str}(a). Note that $\sigma$
is a decreasing function of $\beta$ (or, equivalently, of
temperature). The short distance string tension at the high
temperature side is non--zero due to the magnetic dipoles dynamics
discussed in Sect.~\ref{overview}, Eq.~\eq{sigma:th}.

In order to check the consistency of the results obtained, with the dipole
gas picture we first estimate the dipole size from the string tension
and the monopole density using eq.\eq{sigma:th}.
Taking into account the fact that the density of the dipoles,
$\rho_d$, is half the density of the constituent
monopoles, $\rho$, we get:
\beqn
r = \frac{8 \sigma}{\pi^2 \, \rho}\,.
\label{r}
\eeqn
We evaluate $r$ by this formula and plot it as a function of $\beta$
in Fig.~\ref{figs:r:str}(b). In the deconfinement phase\footnote{In
the confinement phase formula \eq{r} should not work since in this
phase a fraction of the monopoles is not bounded in dipole states.},
$\beta > \beta_c$, the dipole sizes become smaller as $\beta$
increases, in qualitative agreement with the theoretical
estimates. Note that at large $\beta$ the monopole sizes are of the
order of 4 or 5 lattice spacings while the observed sizes are of
the order of
$1, \dots, 2$
spacings.

There are two reasons which may explain observed quantitative
difference in $r$ of a factor of 3. First, the distribution of
the dipole magnetic moment for a real dipole state is unknown while
in our theoretical estimations we have assumed fixed dipole moments
for definiteness. Another reason might be that the analytical
formulae for the dipole density \eq{dens:quant} and the string
tension \eq{sigma:th} are not applicable. This might be caused by
a violation of the dipole diluteness condition \eq{diluteness} or of
the requirement for the density of the dipole moments to be small,
eq.\eq{lambda:mu}.

The quantity characterizing the diluteness of the gas, $\xi$,
eq.\eq{diluteness}, is of the order of $0.1$ for the measured
distances between monopole constituents. However, the dipole sizes
found with the help of eq.\eq{r} give $\xi \sim 0.3$. The last
relation indicates that the point--like dipole gas approach might be
"on the edge of applicability". However, the density of the dipoles as
well as the dipole size decrease with the temperature and therefore
the condition \eq{diluteness} should be satisfied better as
temperatures increases.

The other assumption which has been used in the derivation of the
analytic expressions for the short string tension $\sigma$,
eq.\eq{sigma:th}, and the permittivity $\epsilon$, eq.\eq{Vfull}, is
the smallness of the parameter $\lambda$, eq.\eq{lambda}. Using
Eqs.~(\ref{dens:quant},\ref{sigma:th},\ref{betalat}) we can
express $\lambda$ as a function of the monopole density, the string
tension and the lattice coupling:
\beqn
\lambda = \frac{32 \, \sigma^2 \, \beta}{\pi^2 \, \rho}\,.
\eeqn
With the help of this equation we get that $\lambda$ is a
decreasing function of $\beta$ which is equal to $0.8$ at $\beta =
\beta_c$ and decreases to $0.3$ at $\beta = 2.5$. Thus near the
phase transition and even at the largest studied values of
$\beta$ the requirement \eq{lambda:mu} fails to be fulfilled.
However, at large $\beta$ the coupling $\lambda$ is about to be
small enough for the theoretical formulae for the short string tension
$\sigma$, eq.\eq{sigma:th}, and the permittivity $\epsilon$,
eq.\eq{Vfull}, to start working.

Thus we may conclude that our theoretical formulae may work qualitatively
rather than quantitatively. As we have seen above this is indeed the
case: the linear potential is observed while the mismatch in the
self-consistency check for the dipole sizes for the largest value of
$\beta$ is a factor of $3$. Better agreement between theoretical
and numerical values can be observed for the dielectric permittivity.
Indeed, according to eq.\eq{Vfull}, $\epsilon_{th} - 1 = \lambda
\slash 3 \approx 0.118$ while numerical data shown in
Fig.~\ref{figs:dens:epsilon}(b) gives the result $\epsilon_{num} -
1 = 0.177$.

\section{Conclusions and Acknowledgments}

We have observed the nonperturbative piece in the inter--particle
potential in high temperature compact QED in three dimensions to
rise linearly with the distance. This effect is in
qualitative
agreement with the predictions of the pointlike magnetic
dipole gas model with fixed dipole sizes.
The electric permittivity of the vacuum was also
calculated and turns our to be very close to the theoretical value.

These results may have interesting applications for the physics of
the gauge theories possessing monopole topological excitations which
form bound states. One of these theories is the electroweak model in
which the formation of the Nambu monopole--anti-monopole pairs has
been observed in the high temperature phase~\cite{EW}.

\bigskip
MNCh is grateful to A.~Kovner and V.~I.~Shevchenko for useful
discussions. MNCh and AIV acknowledge the kind hospitality of the
staff of the Department of Physics and Astronomy of the Vrije
University at Amsterdam, where the work was done. Their work was
partially supported by the grants RFBR 99-01-01230a and CRDF award
RP1-2103.

\end{document}